\documentclass[10pt,conference]{IEEEtran}
\usepackage{amsmath,amssymb}
\usepackage{graphicx}
\usepackage{mathrsfs}
\usepackage{tikz}
\usepackage{booktabs}
\usepackage{floatflt}
\usepackage{enumerate}
\usepackage{psfrag}	
\usepackage{array}
\usepackage{multirow,hhline}
\usepackage{exscale}
\usepackage{color}
\usepackage{colortbl}
\usepackage{epsfig,subfigure}
\usepackage{cite}
\usepackage{amsthm}
\usepackage[pdftex,bookmarks=true]{hyperref}

\renewcommand{\vec}[1]{\ensuremath{\mathbf{#1}}}
\newtheorem{theorem}{Theorem}
\newtheorem{lemma}{Lemma}
\newtheorem{remark}{Remark}
\newtheorem{corollary}{Corollary}

\newtheorem{definition}{Definition}

\begin{document}

\IEEEoverridecommandlockouts
\title{On the Sum Capacity of the Y-Channel}
\author{
\IEEEauthorblockN{Anas Chaaban, Aydin Sezgin}
\IEEEauthorblockA{Emmy-Noether Research Group on Wireless Networks\\
Institute of Telecommunications and\\
Applied Information Theory\\
Ulm University, 89081 Ulm, Germany\\
Email: {anas.chaaban@uni-ulm.de, aydin.sezgin@uni-ulm.de}}
\and
\IEEEauthorblockN{Salman Avestimehr}
\IEEEauthorblockA{School of Electrical and Computer Engineering\\
Cornell University\\
325 Frank H.T. Rhodes Hall\\
Ithaca, NY 14853, USA\\
Email: {avestimehr@ece.cornell.edu}}
\thanks{%
The work of A. Chaaban and A. Sezgin is supported by the German Research Foundation, Deutsche
Forschungsgemeinschaft (DFG), Germany, under grant SE 1697/3. The work of A. S. Avestimehr is partly supported by NSF CAREER award  
0953117.%
}
}

\maketitle

\begin{abstract}
A network where three users communicate with each other via a relay is considered. Users do not receive other users' signals via a direct link, and thus the relay is essential for their communication. Each user is assumed to have an individual message to be delivered to each other user. Thus, each user wants to send two messages and to decode two messages. In general, the transmit signals of different nodes can be dependent since they can depend on previously received symbols. We call this case the general case. The sum-capacity is studied, and upper bounds and lower bounds are given. If all nodes have the same power, the sum-capacity is characterized to within a gap of 5/2 bits or a factor of 3 for all values of channel coefficients. This gap is also shown to approach 3/2 bits as the transmit power increases. Moreover, for the symmetric case with equal channel coefficients, the gap is shown to be less than 1 bit. The restricted case is also considered where the transmit signal does not depend on previously received symbols. In this case, the sum-capacity is characterized to within a gap of 2 bits or a factor of 3 for all values of channel coefficients, and approaches 1 bit as the transmit power increases.
\end{abstract}


\section{Introduction}
A multi-way channel is a scenario where users communicate with each other in both directions. The smallest multi-way communication model is the two way channel \cite{Shannon_TWC} where 2 nodes communicate with each other, and each has a message to deliver to the other node. In this sense, each node is a source and a destination at the same time.

The two-way channel can be extended into a bi-directional relay channel by including a relay in the model. In the bi-directional relay channel, two nodes communicate with each other via a relay. This setup was introduced in \cite{RankovWittneben} where relaying protocols were analyzed. In \cite{KimDevroyeMitranTarokh}, further relaying protocols were proposed, and their achievable rate regions were compared to previous work. Achievable schemes for this setup using decode-and-forward and compress-and-forward were studied in \cite{GunduzTuncelNayak} where rate regions were given and capacity was characterized within half a bit for the Gaussian setting. The capacity region of the two-way relay channel was also characterized within a constant gap in \cite{AvestimehrSezginTse}. These results were also extended to the larger network consisting of two pair of nodes in addition to the relay. The approximate capacity of the two-pair bi-directional relay network was obtained in \cite{SezginKhajehnejadAvestimehrHassibi} and \cite{SezginAvestimehrKhajehnejadHassibi}.

If more than two nodes want to communicate via a relay in a bi-directional manner, we get the multi-way relay channel. The multi-way relay channel was studied in \cite{GunduzYenerGoldsmithPoor}, where upper and lower bounds for the capacity of the Gaussian multi-way relay channel were given. In their setup, G\"und\"uz et al. divided users into several clusters, where each user in a cluster has a single message intended to all other users in the same cluster. All users communicate simultaneously via a relay. A similar setup was considered in \cite{OngKellettJohnson}, where all users belong to the same cluster and all channel gains are equal. The authors of \cite{OngKellettJohnson} obtained the sum-capacity of this Gaussian setup with more than 2 users.

In this paper, we consider a Gaussian 3-way relay channel, with a slight difference from the aforementioned multi-way relay channel. In our 3-way channel, 3 users communicate with each other simultaneously via a relay. However, each user has 2 independent messages, each of which is intended to one of the other users. Thus each node wants to broadcast 2 messages to the other nodes, and wants to decode 2 other messages. A MIMO variant of this model was considered in \cite{LeeLim}, where a transmission scheme was proposed, and its corresponding achievable degrees of freedom were calculated. It was referred to as the ``Y-channel".

We consider the single antenna Gaussian case, where all nodes are full-duplex, and derive upper bounds for the sum-capacity of this channel. We distinguish between two cases: a general Y-channel, and a restricted Y-channel. In the general case, the transmit signals of the users can depend on the previously received symbols, while in the restricted case it can not. In addition to the cut-set bounds, new bounds are derived that are shown to be tighter than the cut-set bounds at moderate to high transmit power.

In \cite{OngKellettJohnson}, the so-called ``functional decode and forward" scheme was used as an achievable scheme for the multi-way relay channel. However, in \cite{OngKellettJohnson}, the case where each user has only one message to be delivered to all other users was considered. This is different from our model, where each user has 2 independent message, one for each other user. Thus, we modify the ``functional decode-and-forward" scheme accordingly to obtain a lower bound for the sum-capacity using lattice alignment. Other lower bounds are obtained by using complete decode and forward, or by operating the Y-channel as a bi-directional relay channel where only two users are active at the same time.

Comparing the upper bounds and lower bounds, we bound the gap between them for the case of equal power at all nodes. This gap is shown to be less than 5/2 bits for all values of channel coefficients. Moreover, this gap is shown to approach 3/2 bits as power increases. We also bound the multiplicative gap between the bounds by 3. For the symmetric Y-channel where all channel gains are equal, we show that the gap between these bounds is less than one bit. 

For the restricted Y-channel, the bounds are further tightened, and we characterize the sum-capacity within 2 bits for all values of channel coefficients when all nodes have equal power. This gap is shown to approach 1 bit as power increases.

The rest on the paper is organized as follows. The system model is described in section \ref{Model}. The general Y-channel is considered first, and upper bounds for its sum-capacity are given in section \ref{UpperBounds} and lower bounds in sections \ref{LowerBound1} and \ref{LowerBound2}. The gap between upper and lower bounds is calculated in section \ref{GapCalculation:General}. The restricted Y-channel is considered in section \ref{Section:RYC} and we summarize in section \ref{Summary}. Throughout the paper, we use $x^n$ to denote a sequence of $n$ symbols $(x_1,\dots,x_n)$, we use $C(x)=\frac{1}{2}\log(1+x)$, and $[x]^+=\max\{0,x\}$.

\section{System Model}
\label{Model}

\begin{figure}
\centering
\includegraphics[width=0.8\columnwidth]{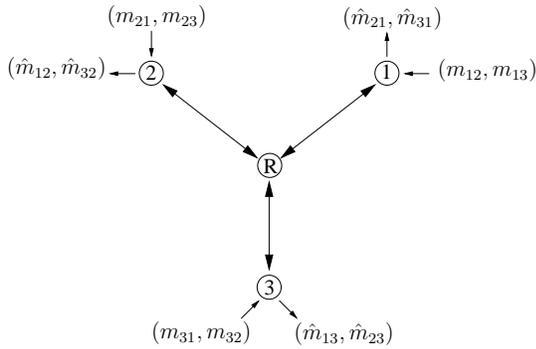}
\caption{The Y-channel: User 1 wants to send two messages, $m_{12}$ to user 2, and $m_{13}$ to user 3. User 1 also wants to decode two messages, $m_{21}$ from user 2, and $m_{31}$ from user 3. Similarly at users 2 and 3.}
\label{Fig:Model}
\end{figure}

The Y-channel models a setup where 3 users want to communicate with each other in a bi-directional manner, and this communication is only possible via a relay as shown in Figure \ref{Fig:Model}. Each user has an individual message to each other users. Consequently, each user wants to broadcast 2 messages via the relay, and wants to decode 2 messages. We assume that all nodes are full duplex, and that there is an AWGN channel between each node and the relay, where the noise is of zero-mean and unit-variance.

User $j$ has messages 
\begin{align}
m_{jk}&\in\mathcal{M}_{jk}\triangleq\{1,\dots,2^{nR_{jk}}\}, \text{and }\\ m_{jl}&\in\mathcal{M}_{jl}\triangleq\{1,\dots,2^{nR_{jl}}\}
\end{align}
to users $k$ and $l$ respectively where $R_{jk},R_{jl}\in\mathbb{R}_+$, for all distinct $j,k,l\in\{1,2,3\}$. The messages of user $j$ are encoded into a sequence $x_j^n$ using an encoder $f_j$, where for $i=1,\dots,n$, $x_{ji}$ is a realization of a real random variable $X_{ji}$ such that 
\begin{align}
\frac{1}{n}\sum_{i=1}^n\mathbb{E}[X_{ji}^2]\leq P.
\end{align}
The codeword $x_j^n$ can be generated in different ways according to the following cases \cite{Shannon_TWC}:
\begin{itemize}
\item[1)] General encoding: $x_j^n$ is a function of $m_{jk}$, $m_{jl}$, and the previously received symbols at node $j$, thus
\begin{align}
\label{SourceEncoder}
x_{ji}=f_j(m_{jk},m_{jl},y_j^{i-1}).
\end{align}
\item[2)]Restricted encoding: $x_j^n$ is a function of $m_{jk}$ and $m_{jl}$ only, thus
\begin{align}
\label{SourceEncoder_R}
x_{j}^n=f_j(m_{jk},m_{jl}).
\end{align}
\end{itemize}
In the Y-channel with general encoding, which we call a general Y-channel, the transmit signals of different users are dependent. This is not the case with restricted encoding in what we call a restricted Y-channel. 

The received signal at the relay at time instant $i$ can be written as
\begin{align}
y_{ri}=h_1x_{1i}+h_2x_{2i}+h_3x_{3i}+z_{ri},
\end{align}
where $z_{ri}$ is a realization of an i.i.d. Gaussian noise $Z_r\sim\mathcal{N}(0,1)$ and $h_1,h_2,h_3\in\mathbb{R}$ are the channel coefficients from the users to the relay. We assume without loss of generality that 
\begin{align}
\label{Ordering}
h_1^2\geq h_2^2\geq h_3^2.
\end{align}
The relay sends a sequence $x_r^n$ of random variables $X_{ri}$ that satisfy 
\begin{align}
\frac{1}{n}\sum_{i=1}^n\mathbb{E}[X_{ri}^2]\leq P_r,
\end{align}
which depends on the past received symbols at the relay, i.e.
\begin{align}
\label{RelayEncoder}
X_{ri}=f_r(Y_r^{i-1}).
\end{align}
Then, the received signal at user $j$ and time $i$ can be written as
\begin{align}
\label{ReceivedSignal}
y_{ji}=h_jx_{ri}+z_{ji},
\end{align}
where $z_{ji}$ is a realization of an i.i.d. Gaussian noise $Z_j\sim\mathcal{N}(0,1)$. We have assumed that the channel is reciprocal, i.e. the channel gain from user $j$ to the relay is the same as that from the rely to user $j$. Each node $j$ uses a decoding function $g_j$ to decode $m_{kj}$ and $m_{lj}$, i.e.
\begin{align}
(\hat{m}_{kj},\hat{m}_{lj})=g_j(y_j^n,m_{jk},m_{jl}).
\end{align}

\begin{definition}
We denote the vector of all rates by $\vec{R}$
and that of all messages by $\vec{m}$
\begin{align}
\vec{R}&=(R_{12},R_{13},R_{21},R_{23},R_{31},R_{32})\\
\vec{m}&=(m_{12},m_{13},m_{21},m_{23},m_{31},m_{32})
\end{align}
We also define $R_\Sigma(\vec{R})$ to be the sum of the components of $\vec{R}$ or
\begin{align}
R_\Sigma(\vec{R})=\sum_{j=1}^3\sum_{\substack{k=1\\ k\neq j}}^3R_{jk},
\end{align}
\end{definition}

The message sets $\mathcal{M}_{jk}$, encoding functions $f_j$, $f_r$, and decoding functions $g_j$ define a code $(\vec{R},n)$ for the Y-channel. An error occurs if $(\hat{m}_{kj},\hat{m}_{lj})\neq({m}_{kj},{m}_{lj})$, for distinct $j,k,l\in\{1,2,3\}$. A rate tuple $\vec{R}\in\mathbb{R}_+^6$ is achievable if there exist a sequence of $(\vec{R},n)$ codes with an average error probability that approaches zero as $n$ increases. The set of all achievable rate tuples is the capacity region $\mathcal{C}$ of the Y-channel. An achievable sum-rate is $R_\Sigma(\vec{R})$ where $\vec{R}\in\mathcal{C}$ or simply $R_\Sigma$
and the sum-capacity is the maximum achievable sum rate given by
\begin{align}
C=\max_{\vec{R}\in\mathcal{C}}R_\Sigma.
\end{align}

In the following sections, we will deal with the sum-capacity of the Y-channel, by deriving upper and lower bounds. Then we bound the gap between the upper and lower bounds. We consider both the general Y-channel where the encoding functions are as given in (\ref{SourceEncoder}) whose sum-capacity will be denoted $C_g$, and the restricted Y-channel where the encoding functions are as given in (\ref{SourceEncoder_R}) whose sum-capacity will be denoted $C_r$. Clearly, $C_r\leq C_g$.

\section{General Y-channel: Upper bounds}
\label{UpperBounds}
\begin{figure}[t]
\centering
\includegraphics[width=0.8\columnwidth]{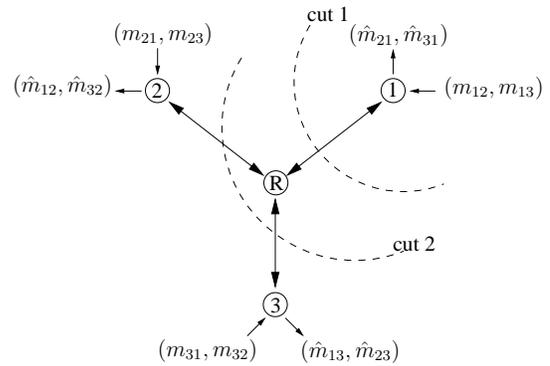}
\caption{A cut in the Y-channel. Cut 1 splits the set $\mathcal{S}=\{U_1,U_2,U_3,R\}$ into $\mathcal{T}=\{U_1\}$ and $\mathcal{T}^c$. This can be used to obtain a bound on $R_{12}+R_{13}$ if we consider information flow from $\mathcal{T}$ to $\mathcal{T}^c$, and on $R_{21}+R_{31}$ if we consider information flow from $\mathcal{T}^c$ to $\mathcal{T}$. Similarly, using cut 2 we can obtain one more bound on both $R_{12}+R_{13}$ and $R_{21}+R_{31}$.}
\label{Cut}
\end{figure}

We start by considering the general Y-channel, and give sum-capacity upper bounds for this case. One way to obtain upper bounds for the Y-channel is by using the cut-set bounds \cite{CoverThomas}. If we label the set of nodes in the Y-channel by $\mathcal{S}\triangleq\{U_1,U_2,U_3,R\}$ where $U_j$ denotes user $j$, $j\in\{1,2,3\}$ and $R$ denotes the relay, then the cut-set bounds provide upper bounds on the rate of information flow from a set $\mathcal{T}\subset\mathcal{S}$ to its complement $\mathcal{T}^c$ in $\mathcal{S}$ (see Figure \ref{Cut}). The cut-set bounds for this setup yield the following upper bounds.

\begin{theorem}
\label{CutSetBounds}
The achievable rates in the Y-channel are upper bounded by
\begin{align}
\label{CS1}
&\hspace{-0.3cm}R_{jk}+R_{jl}\leq\min\left\{I(X_j;Y_r|X_k,X_l,X_r),\right.\nonumber\\
&\hspace{4.5cm}\left.I(X_r;Y_k,Y_l|X_k,X_l)\right\}\\
\label{CS2}
&\hspace{-0.3cm}R_{jl}+R_{kl}\leq\min\left\{I(X_j,X_k;Y_r|X_l,X_r),I(X_r;Y_l|X_l)\right\}
\end{align}
for all distinct $j,k,l\in\{1,2,3\}$, where $(X_1,X_2,X_3,X_r)$ is a zero-mean Gaussian random vector with joint distribution $p(x_1,x_2,x_3,x_r)$, such that $\mathbb{E}[X_j^2]\leq P$ and $\mathbb{E}[X_r^2]\leq P_r$.
\end{theorem}

The first bound (\ref{CS1}) in Theorem \ref{CutSetBounds} is obtained by considering the cuts $\mathcal{T}=\{U_j\}$ and $\mathcal{T}=\{U_j,R\}$, respectively for the first and second arguments of the $\min$ operation. These cuts are shown for the case of $j=1$ in Figure \ref{Cut} labeled as cut 1 and cut 2 respectively. The last bound (\ref{CS2}) in Theorem \ref{CutSetBounds} is obtained by considering the complementary cuts. Namely, the first and the second arguments of the $\min$ operation are obtained by considering $\mathcal{T}=\{U_j,R\}^c$ and $\mathcal{T}=\{U_j\}^c$ respectively. The following bounds are obtained as a corollary from Theorem \ref{CutSetBounds}.

\begin{corollary}
\label{CSG}
The achievable rates in the Y-channel must satisfy
\begin{align}
\label{CSG1}
R_{jk}+R_{jl}&\leq C\left(\min\left\{h_j^2P,h_k^2P_r+h_l^2P_r\right\}\right)\\
\label{CSG2}
R_{jl}+R_{kl}&\leq C\left(\min\left\{(|h_j|+|h_k|)^2P,h_l^2P_r\right\}\right),
\end{align}
for all distinct $j,k,l\in\{1,2,3\}$.
\end{corollary}
\begin{proof}
See Appendix \ref{CSGProof}.
\end{proof}

In the following theorem, we give other bounds on the achievable rates in the Y-channel based on a degraded broadcast channel bound.

\begin{figure}[t]
\centering
\includegraphics[width=0.8\columnwidth]{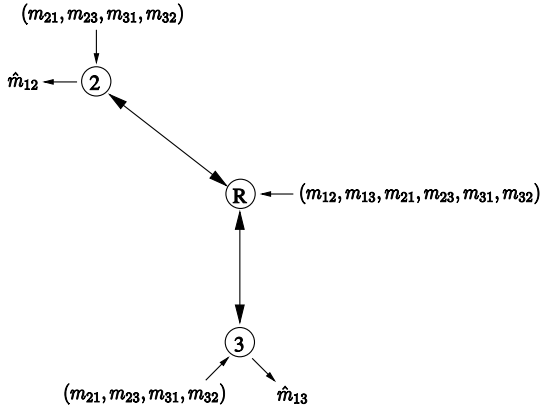}
\caption{Genie aided Y-Channel as a degraded broadcast channel.}
\label{GA_BC}
\end{figure}

\begin{theorem}
\label{BCBounds}
The achievable rates in the Y-channel must satisfy
\begin{align}
R_{12}+R_{13}&\leq C(h_2^2P_r),\\
R_{21}+R_{23}&\leq C(h_1^2P_r),\\
R_{31}+R_{32}&\leq C(h_1^2P_r).
\end{align}
\end{theorem}
\begin{proof}
Let us give the relay all the messages as side information, i.e. the relay knows $\vec{m}$ apriori. And let us also give $(m_{31},m_{32})$ and $(m_{21},m_{23})$ as side information to receivers 2 and 3 respectively (see Figure \ref{GA_BC}). Now receivers 2 and 3 share the knowledge of $m_{21}$, $m_{23}$, $m_{31}$ and $m_{32}$ which are also known at the relay. The relay knows $m_{12}$ and $m_{13}$ which should be delivered to receivers 2 and 3 respectively. The resulting setup is a degraded broadcast channel (BC) whose sum-capacity is \cite{CoverThomas}
\begin{align}
R_{12}+R_{13}\leq C(\max\{h_2^2,h_3^2\}P_r).
\end{align}
Similarly, we can obtain bounds on $R_{21}+R_{23}$ and $R_{31}+R_{32}$. Using (\ref{Ordering}), we obtain the statement of the theorem.
\end{proof}

The bounds in Corollary \ref{CSG} and Theorem \ref{BCBounds}, in addition to $R_{jk}\geq0$ and the single user bounds
\begin{align}
R_{jk}\leq \min\{C(h_j^2P),C(h_k^2P_r)\}
\end{align}
for all distinct $j,k\in\{1,2,3\}$ form a region $\overline{\mathcal{C}}$ in the 6-dimensional space which is an outer bound on the capacity region $\mathcal{C}$ of the Y-channel. In order to find an upper bound on the sum capacity $C_g$, we have to solve
\begin{align}
\max_{\vec{R}\in\overline{\mathcal{C}}} R_\Sigma,
\end{align}
or otherwise, we can add any three bounds from Corollary \ref{CSG} and Theorem \ref{BCBounds} whose left hand side terms add to $R_\Sigma$. However, such an upper bound will depend on the relative value of $P_r$ compared to $P$. If we specialize these results to the Y-channel with $P=P_r$ we get a simpler representation for a sum-capacity upper bound. By combining the bounds in Corollary \ref{CSG} and the bounds in Theorem \ref{BCBounds}, we obtain the following corollary.
\begin{corollary}
\label{Corollary:CutSetBounds}
If $P=P_r$, then the sum-capacity of the Y-channel is upper bounded by $\overline{C}_{\Sigma}$, i.e. 
\begin{align}
C_g\leq\overline{C}_{\Sigma}&\triangleq 2C(h_2^2P)+C( h_3^2P).
\end{align}
\end{corollary}
\begin{proof}
By evaluating the bounds in corollary \ref{CSG} for $P=P_r$, we have
\begin{align}
\label{CSBnd1}
R_{12}+R_{13}&\leq C(\min\{h_1^2,h_2^2+h_3^2\}P),\\
\label{CSBnd2}
R_{21}+R_{23}&\leq C(\min\{h_2^2,h_1^2+h_3^2\}P),\\
\label{CSBnd3}
R_{31}+R_{32}&\leq C(\min\{h_3^2,h_1^2+h_2^2\}P).
\end{align}
Moreover, from Theorem \ref{BCBounds}, we have
\begin{align}
\label{BCBnd1}
R_{12}+R_{13}&\leq C(h_2^2P),
\end{align}
if $P=P_r$, which is more binding than (\ref{CSBnd1}) due to (\ref{Ordering}). Adding (\ref{CSBnd2}), (\ref{CSBnd3}), and (\ref{BCBnd1}) and using (\ref{Ordering}) we obtain
\begin{align}
R_\Sigma&\leq C(h_2^2P)+C(h_2^2P)+C(h_3^2P)
\end{align}
and the statement of the corollary follows.
\end{proof}
\begin{remark}
The upper bound in Corollary \ref{Corollary:CutSetBounds} is independent of $h_1$ due to the assumption in (\ref{Ordering}).
\end{remark}

In \cite{GunduzYenerGoldsmithPoor} and \cite{OngKellettJohnson}, the multi-cast setting of the multi-way relay channel was considered, where each node has one message intended to all other nodes. In that case, it was shown that the cut-set bounds are sufficient to obtain an asymptotic characterization of the sum-capacity. Interestingly however, in our broadcast setting this is not the case. We can notice that the bound of Corollary \ref{Corollary:CutSetBounds}, which is based on the cut-set bounds in Theorem \ref{CutSetBounds}, provides a sum-capacity upper bound of the form 
\begin{align*}
C_g\leq\frac{3}{2}\log(P)+o(\log(P)).
\end{align*}
Thus, this corollary gives a sum-capacity upper bound with a pre-log of 3/2. The reason behind this is that Theorems \ref{CutSetBounds} and \ref{BCBounds} bound the sum of two rates by $\frac{1}{2}\log(P)+o(\log(P))$. Next, we develop more upper bounds, and show that the Y-channel has a sum-capacity pre-log of 1,
\begin{align}
C_g\leq\log(P)+o(\log(P)).
\end{align}
This means that, while the bound in Corollary \ref{Corollary:CutSetBounds} might be useful at lower $P$, it can not give a tight sum-capacity upper bound as $P$ increases. Thus, contrary to the multi-cast setting, the cut-set bounds are not sufficient in the broadcast setting and more bounds are required for an asymptotic characterization of the sum-capacity. A bounds with a capacity pre-log of 1 is given in Theorem \ref{SRUBG}. Before we state this theorem, we need the following lemmas.

\begin{lemma}
\label{FromRelay}
The achievable rates in the Y-channel 
must satisfy
\begin{align}
R_{kj}+R_{lj}+R_{kl}&\leq C(h_j^2P_r+h_l^2P_r)
\end{align}
for all distinct $j,k,l\in\{1,2,3\}$.
\end{lemma}
\begin{proof}
We use a genie aided approach to bound the sum of three rates, e.g. $R_{21}+R_{31}+R_{23}$, by giving $m_{32}$ and $(Y_1^n,m_{21},m_{12},m_{13})$ as additional information to receivers 1 and 3 respectively. Details are given in Appendix \ref{ProofFromRelay}.
\end{proof}

\begin{lemma}
\label{GUB}
The achievable rates in the Y-channel must satisfy
\begin{align}
R_{kj}+R_{lj}+R_{kl}&\leq C((|h_k|+|h_l|)^2P)
\end{align}
for all distinct $j,k,l\in\{1,2,3\}$.
\end{lemma}
\begin{proof}
We use a genie aided approach to bound the sum of three rates such as $R_{21}+R_{31}+R_{23}$ by giving $(Y_r^n,m_{32})$ and $(Y_r^n,m_{21},m_{12},m_{13})$ as additional information to receiver 1 and 3 respectively. See Appendix \ref{GeneralProof} for more details.
\end{proof}

As a result of Lemmas \ref{FromRelay} and \ref{GUB}, we obtain 
\begin{align}
&R_{kj}+R_{lj}+R_{kl}\nonumber\\
\label{3BoundG}
&\hspace{0.5cm}\leq \min\left\{C(h_j^2P_r+h_l^2P_r),C((|h_k|+|h_l|)^2P)\right\}.
\end{align}
Now if the Y-channel has $P=P_r$, we obtain the following sum-capacity upper bound.

\begin{theorem}
\label{SRUBG}
The sum-capacity of the Y-channel with $P=P_r$ is upper bounded by $\overline{C}_{\Sigma g}$, i.e.
\begin{align}
C_g\leq\overline{C}_{\Sigma g}&=C(h_2^2P+h_3^2P)\nonumber\\
&\quad+C(\min\{h_1^2P+h_3^2P,(|h_2|+|h_3|)^2P\}).
\end{align}
\end{theorem}
\begin{proof}
By evaluating (\ref{3BoundG}) for $(j,k,l)=(2,1,3)$, and for $(j,k,l)=(1,3,2)$ and  adding the two obtained bounds, we obtain the desired result.
\end{proof}

As we can see, the bound in Theorem \ref{SRUBG} has a pre-log equal to 1. The slope of this bound is lower than that of the bound $\overline{C}_{\Sigma}$ obtained with the cut-set approach,  which makes it tighter as $P$ increases.

Next, we provide achievability schemes for the Y-channel where we use complete decode-and-forward, and functional decode-and-forward.

\section{Lower Bound: Complete Decode and Forward}
\label{LowerBound1}
We describe a complete decode and forward scheme for the Y-channel. In this scheme, user $j$ encodes his messages $m_{jk}$ and $m_{jl}$ into an i.i.d. sequence $x_j^n(m_{jk},m_{jl})$ where $X_j\sim\mathcal{N}(0,P)$. Then, all users transmit their signals to the relay together. The relay decodes all messages  in a MAC fashion, with a small probability of error if the rate tuple $\vec{R}$ is in the capacity region from the 3 users to the relay. Hence, we get the following sum-rate constraint in the uplink
\begin{align}
\label{MACconstraint}
R_\Sigma\leq C(h_1^2P+h_2^2P+h_3^2P).
\end{align}

The relay decodes $\vec{m}$ from its received signal. Then it uses a Gaussian codebook to encode $\vec{m}$ into an i.i.d. sequence $x_r^n(\vec{m})$ where $X_r\sim\mathcal{N}(0,P_r)$. The relay then sends $x_r^n(\vec{m})$. After receiving a noisy observation of $x_r^n(\vec{m})$, user 1 knowing $m_{12}$ and $m_{13}$ can decode all other messages as long as \cite{GunduzYenerGoldsmithPoor}
\begin{align}
\label{DFC1}
R_{21}+R_{23}+R_{31}+R_{32}\leq C(h_1^2P_r).
\end{align}
Similarly at the other receivers, reliable decoding is guaranteed if the following rate constraints are fulfilled
\begin{align}
R_{12}+R_{13}+R_{31}+R_{32}&\leq C(h_2^2P_r)\\
\label{DFC2}
R_{12}+R_{13}+R_{21}+R_{23}&\leq C(h_3^2P_r).
\end{align}
In order to find the maximum achievable sum rate, we solve
\begin{align}
\label{OptProb}
\text{maximize}\quad &\sum_{j=1}^3\sum_{\substack{k=1\\k\neq j}}R_{jk}\\
\text{subject to}\quad & R_{jk}\geq0 \quad\forall j,k\in\{1,2,3\},\ j\neq k\nonumber\\
& R_{21}+R_{23}+R_{31}+R_{32}\leq C(h_1^2P_r)\nonumber\\
&R_{12}+R_{13}+R_{31}+R_{32}\leq C(h_2^2P_r)\nonumber\\
&R_{12}+R_{13}+R_{21}+R_{23}\leq C(h_3^2P_r).\nonumber
\end{align}
Solving the linear program (\ref{OptProb}) keeping (\ref{Ordering}) in mind, we obtain (see Appendix \ref{LPMax})
\begin{align}
\label{BCconstraints}
R_\Sigma\leq \min\left\{\sum_{j=2}^3C(h_j^2P_r),\frac{1}{2}\sum_{j=1}^3C(h_j^2P_r)\right\}.
\end{align}

Hence, we obtain the following lower bound for the sum-capacity.
\begin{theorem}
\label{LowerBound:DF}
The sum-capacity of the Y-channel satisfies $C_g\geq\underline{C}^I$ where
\begin{align}
\label{DFconstraint}
\underline{C}^I&=\min\left\{C\left(\sum_{j=1}^3h_j^2P\right),\sum_{j=2}^3C(h_j^2P_r),\frac{1}{2}\sum_{j=1}^3C(h_j^2P_r)\right\}.
\end{align}
\end{theorem}
\begin{proof}
The maximum achievable sum rate using complete decode and forward is given by the minimum of (\ref{MACconstraint}) and (\ref{BCconstraints}). Therefore (\ref{DFconstraint}) is a lower bound for the sum-capacity.
\end{proof}

\section{Lower Bound: Functional Decode and Forward}
\label{LowerBound2}
In this section, we describe another achievable scheme that gives us a lower bound for the sum-capacity of the Y-channel. In this scheme, time is divided into  frames of 3 time slots each, where in each slot, only 2 users and the relay are active. These blocks will be indexed as block $3f+s$ where $f\in\mathbb{N}$ denotes the frame index and $s\in\{1,2,3\}$ the slot index.

Briefly, in block $3f+s$, the two active users send, say $x_1^n(m_{12}(f))$ and $x_2^n(m_{21}(f))$ to the relay, $m_{12}(f),m_{21}(f)\in\{1,\dots,2^{nR_{12}}\}$. The relay decodes the superposition of $x_1^n(m_{12}(f))$ and $x_2^n(m_{21}(f))$ (in a way that will be explained next), maps it to $u_{12}(f)\in\{1,\dots,2^{nR_{12}}\}$ and sends $x_r^n(u_{12}(f))$ in block $3f+s+1$. Table \ref{Cycle} illustrates the 3 main blocks used.
\begin{table*}[ht]
\centering
\begin{tabular}{|c||c|c|c|c|c|}
\hline
Block & Node & 1 & 2 & 3 & relay\\\hline
\multirow{2}{*}{4} & sends & $m_{12}(1)$ & $m_{21}(1)$ & - & $u_{31}(0)$\\
& decodes & $m_{31}(0)$ & - & $m_{13}(0)$ & $X_1^n(m_{12}(1))+X_2^n(m_{21}(1))\to u_{12}(1)$\\\hline
\multirow{2}{*}{5} & sends & - & $m_{23}(1)$ & $m_{32}(1)$ & $u_{12}(1)$\\
& decodes & $m_{21}(1)$ & $m_{12}(1)$ & - & $X_2^n(m_{23}(1))+X_3^n(m_{32}(1))\to u_{23}(1)$\\\hline
\multirow{2}{*}{6} & sends & $m_{13}(1)$ & - & $m_{31}(1)$ & $u_{23}(1)$\\
& decodes & - & $m_{32}(1)$ & $m_{23}(1)$ & $X_1^n(m_{13}(1))+X_3^n(m_{31}(1))\to u_{31}(1)$\\\hline
\end{tabular}
\caption{Three transmission blocks of the achievability scheme shown for frame $f=1$. }
\label{Cycle}
\end{table*}

These three blocks are of length $n$ symbols each. The procedure in block $3f+s$ is the same as that in block $s$. Notice that each user transmits in only 2 out of 3 slots. In what follows, we illustrate the scheme for blocks $3f+1$, $3f+2$, and $3f+3$. We remove the frame index from the messages for readability.

\subsection{Codebook generation}
The users use nested lattice codebooks. We start with some lattice preliminaries. An $n$-dimensional lattice $\Lambda$ is a subset of $\mathbb{R}^n$ such that $\lambda_1,\lambda_2\in\Lambda\Rightarrow\lambda_1+\lambda_2\in\Lambda$, i.e. it is an additive subgroup of $\mathbb{R}^n$. The fundamental Voronoi region $\mathcal{V}(\Lambda)$ of $\Lambda$ is the set of all points in $\mathbb{R}^n$ whose distance to the origin is smaller that that to any other $\lambda\in\Lambda$. Thus, by quantizing points in $\mathbb{R}^n$ to their closest lattice point, all points in $\mathcal{V}(\Lambda)$ are mapped to the all zero vector.

Two lattices are considered for nested lattice codes, a coarse lattice $\Lambda_c$ and a fine lattice $\Lambda_f$ where $\Lambda_c\subseteq\Lambda_f$. The codewords are chosen as the fine lattice points $\lambda_f\in\Lambda_f$ that lie in $\mathcal{V}(\Lambda_c)$. The power constraint is satisfied by an appropriate choice of $\Lambda_c$ and the rate of the code is defined by the number of fine lattice points in $\{\Lambda_f\cap\mathcal{V}(\Lambda_c)\}$ (codewords).

We denote the lattice corresponding to the message set $\mathcal{M}_{jk}$ by $\Lambda_{jk}$ with rate $R_{jk}$. Furthermore, we fix the rates such that $R_{jk}=R_{kj}$. Each message $m_{jk}$ is mapped into a codeword (lattice point) $x_j^n(m_{jk})=v_{jk}\in\Lambda_{jk}$. The lattices are constructed in such a way that the following alignment equations are satisfied:
\begin{align}
h_1\Lambda_{12}&=h_2\Lambda_{21}\\
h_1\Lambda_{13}&=h_3\Lambda_{31}\\
h_2\Lambda_{23}&=h_3\Lambda_{32}
\end{align}

The relay uses three Gaussian codebooks of rate $R_{12}$, $R_{23}$, and $R_{31}$. That is, e.g. it generates $2^{nR_{12}}$ i.i.d sequences $X_r^n$ where $X_r\sim\mathcal{N}(0,P_r)$. Each sequence is given an index $u_{12}\in\mathcal{U}_{12}\triangleq\{1,\dots,2^{nR_{12}}\}$. In this scheme, the relay communicates with two users at a time, we use $u_{ij}$ to indicate that the message sent carries information to both users $i$ and $j$.

\subsection{Encoding at the sources}
The encoding at the sources in block $3f+1$ is done as follows. Users 1 and 2 map $m_{12}$ and $m_{21}$ to codewords (lattice points) $x_1^n(m_{12})=v_{12}$ and $x_2^n(m_{21})=v_{21}$ respectively, with $v_{12}\in\Lambda_{12}$ and $v_{21}\in\Lambda_{21}$. Then they transmit these codewords. Users 2 and 3 transmit $x_2^n(m_{23})$ and $x_3^n(m_{32})$ respectively in block $3f+2$, and users 3 and 1 transmit $x_3^n(m_{31})$ and $x_1^n(m_{13})$ respectively in block $3f+3$.

\subsection{Processing at the relay}
The received signal at the relay in block $3f+1$ is 
\begin{align}
y_r^n&=h_1x_1^n+h_2x_2^n+z_r^n\nonumber\\
&=h_1v_{12}+h_2v_{21}+z_r^n.
\end{align}
Notice that $h_1v_{12}+h_2v_{21}$ is also a lattice point $h_1v_{12}+h_2v_{21}\in h_1\Lambda_{12}$. The relay can decode the superposition $h_1v_{12}+h_2v_{21}$ with arbitrarily small probability of error if \cite{NarayananPravinSprintson,NamChungLee}
\begin{align}
R_{12}&=R_{21}\leq\left[C\left(h_1^2P'-\frac{1}{2}\right)\right]^+\\
R_{12}&=R_{21}\leq\left[C\left(h_2^2P'-\frac{1}{2}\right)\right]^+,
\end{align}
where $P'$ is the transmit power. Since each user transmits in 2 blocks out of 3, we can set $P'=3P/2$ without violating the power constraint of the users. Thus, the following rates are achievable
\begin{align}
R_{12}=R_{21}\leq\left[C\left(\frac{3h_2^2P}{2}-\frac{1}{2}\right)\right]^+,
\end{align}
since $h_2\leq h_1$. At the end of block $3f+1$, the relay knows $h_1v_{12}+h_2v_{21}\in h_1\Lambda_{12}$, and maps it to an index $u_{12}\in\mathcal{U}_{12}$. Then, it maps $u_{12}$ into a codeword $x_r^n(u_{12})$, and transmits $x_r^n(u_{12})$ in the next block, block $3f+2$\footnote{At the beginning of transmission, the relay does not send anything. This results in a loss in the achievable rate. However, this loss becomes negligible as $b$ increases.}. Keep in mind that this message $u_{12}$ is meant for users 1 and 2. 

In block $3f+2$, the relay decodes $h_2v_{23}+h_3v_{32}$, maps it to $u_{23}\in\mathcal{U}_{23}$ and sends $x_r^n(u_{23})$, and in block $3f+3$ the relay decodes $h_1v_{13}+h_3v_{31}$, maps it to $u_{31}\in\mathcal{U}_{31}$ and sends $x_r^n(u_{31})$ (cf. Table \ref{Cycle}).

\subsection{Decoding at the destinations}
At the end of the block $3f+1$, the first and third users have $y_1^n=h_1x_r^n+z_1^n$ and $y_3^n=h_3x_r^n+z_3^n$ and aim to decode $u_{31}$. This can be done with an arbitrarily small probability of error if 
\begin{align}
R_{31}=R_{13}&\leq C\left(h_1^2P_r\right)\\
R_{13}=R_{13}&\leq C\left(h_3^2P_r\right).
\end{align}
Knowing $u_{31}$, users 1 and 3 are able to calculate $h_1v_{13}+h_3v_{31}$ and since each knows his own message $m_{13}$ and $m_{31}$ respectively, they can obtain $m_{31}$ and $m_{13}$. Similarly, users 1 and 2 decode $m_{21}$ and $m_{12}$ in block $3f+2$, and users 2 and 3 decode $m_{32}$ and $m_{23}$ in block $3f+3$.

As a result, the achievable rate using this scheme is bounded by
\begin{align*}
R_{12}&\leq\min\left\{\left[C\left(\frac{3h_2^2P}{2}-\frac{1}{2}\right)\right]^+,C\left(h_2^2P_r\right)\right\}\\
R_{13}&\leq\min\left\{\left[C\left(\frac{3h_3^2P}{2}-\frac{1}{2}\right)\right]^+,C\left(h_3^2P_r\right)\right\}\\
R_{23}&\leq\min\left\{\left[C\left(\frac{3h_3^2P}{2}-\frac{1}{2}\right)\right]^+,C\left(h_3^2P_r\right)\right\},
\end{align*}
and $R_{12}=R_{21}$, $R_{13}=R_{31}$, $R_{23}=R_{32}$. Since we have used 3 blocks to transmit all messages, we obtain the following theorem.
\begin{theorem}
\label{LowerBound:FDF}
The sum-capacity of the Y-channel satisfies $C_g\geq\underline{C}^{II}$ where
\begin{align*}
\underline{C}^{II}&=\frac{2}{3}\min\left\{\left[C\left(\frac{3h_2^2P}{2}-\frac{1}{2}\right)\right]^+,C\left(h_2^2P_r\right)\right\}\\
&+\frac{4}{3}\min\left\{\left[C\left(\frac{3h_3^2P}{2}-\frac{1}{2}\right)\right]^+,C\left(h_3^2P_r\right)\right\}.
\end{align*}
\end{theorem}

\subsection{Functional decode and forward with two active users}
We can also obtain a sum-capacity lower bound by letting two out of three users communicate all the time as in a two-way relay channel. By choosing the strongest two users to communicate all the time, i.e. users 1 and 2, the following rates can be achieved \cite{NamChungLee}
\begin{align}
R_{12}=R_{21}\leq\min\left\{\left[C\left(h_2^2P-\frac{1}{2}\right)\right]^+,C(h_2^2P_r)\right\}
\end{align}
Thus, we can bound the sum-capacity as follows.
\begin{theorem}
\label{LowerBound:FDF_2User}
The sum-capacity of the Y-channel satisfies $C_g\geq\underline{C}^{III}$ where
\begin{align*}
\underline{C}^{III}&=2\min\left\{\left[C\left(h_2^2P-\frac{1}{2}\right)\right]^+,C\left(h_2^2P_r\right)\right\}.
\end{align*}
\end{theorem}

\begin{figure}[t]
\centering
\psfragscanon
\psfrag{x}[t]{$\mathsf{SNR}$(dB)}
\psfrag{y}[b]{Sum Rate (Bits/channel use)}
\psfrag{CS}[l]{\scriptsize {Upper bound: $\overline{C}_{\Sigma}$}}
\psfrag{BR}[l]{\scriptsize {Upper bound: $\overline{C}_{\Sigma r}$}}
\psfrag{BG}[l]{\scriptsize {Upper bound: $\overline{C}_{\Sigma g}$}}
\psfrag{DF}[l]{\scriptsize {Lower bound: $\underline{C}^I$}}
\psfrag{FF}[l]{\scriptsize {Lower bound: $\underline{C}^{II}$}}
\psfrag{2W}[l]{\scriptsize {Lower bound: $\underline{C}^{III}$}}
\includegraphics[width=\columnwidth]{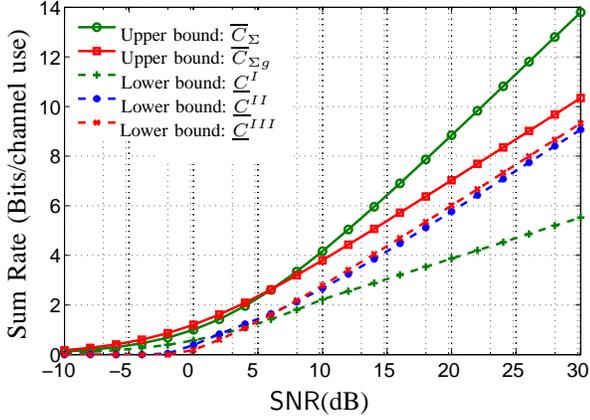}
\caption{A plot of the upper and lower bounds for a Y-channel with $P=P_r$, $h_1=1$, $h_2=0.8$, and $h_3=0.7$.}
\label{sum_rate_asymmetric}
\end{figure}

Figure \ref{sum_rate_asymmetric} shows a plot of the obtained upper and lower bounds for the case $P=P_r$ versus the signal to noise power ratio $\mathsf{SNR}$. Namely, the plotted bounds are: the upper bound obtained with the cut-set approach $\overline{C}_{\Sigma}$, the upper bound obtained with the genie aided approach $\overline{C}_{\Sigma g}$, the complete decode-and-forward lower bound $\underline{C}^I$, the functional decode-and-forward lower bound $\underline{C}^{II}$, and the functional decode-and-forward lower bound with two active users $\underline{C}^{III}$, for a Y-channel with $h_1=1$, $h_2=0.8$, $h_3=0.7$. It can be seen that $\overline{C}_{\Sigma g}$ is tighter than $\overline{C}_{\Sigma}$ at moderate to high $\mathsf{SNR}$. It can also be seen that the gap between $\overline{C}_{\Sigma g}$ and $\underline{C}^{II}$, $\underline{C}^{III}$ becomes constant as $\mathsf{SNR}$ increases. In the following section, we characterize this constant gap. Notice that the lower bound $\underline{C}^{III}$ is simpler than $\underline{C}^{II}$. For this reason, we will use $\underline{C}^{III}$ to characterize that gap between the upper and lower bounds. However, it must be noted that $\underline{C}^{II}$ can be larger than $\underline{C}^{III}$ in some cases, e.g. if $h_3=h_2$.

\section{Bounding the Gap between the Upper and Lower Bounds}
\label{GapCalculation:General}

The functional decode and forward scheme achieves the DoF of the Y-channel. This can be seen from the pre-log in the lower bound in Theorem \ref{LowerBound:FDF} and the upper bounds in Theorem \ref{SRUBG}. Now we bound the gap between the upper and lower bounds. We consider two kinds of gaps, additive gap and multiplicative gap.

We bound the multiplicative gap $\Gamma_m$ first. That is, we bound the ratio of the upper bound to the lower bound. For this purpose, we use the bounds $\overline{C}_\Sigma$ and $\underline{C}^I$. Notice that we can always write
\begin{align}
C_g&\geq\underline{C}^{I}\geq C\left(h_2^2P\right).
\end{align}
Therefore
\begin{align}
\Gamma_{m}&=\frac{\overline{C}_\Sigma}{\underline{C}^I}\\
&\leq\frac{3C(h_2^2P)}{C(h_2^2P)}\\
&\leq 3.
\end{align}

Now we calculate the additive gap, which we split into two cases: $h_2^2P\leq1/2$ and $h_2^2P>1/2$.

\subsection{Case $h_2^2P\leq1/2$}
In this case, we call the gap $\Gamma_{a1}$. Consider the lower bound $\underline{C}^I$ and the upper bound $\overline{C}_\Sigma$. These bounds can be used to obtain the following.
\begin{align}
\Gamma_{a1}&=\overline{C}_\Sigma-\underline{C}^{I}\\
&\leq C(h_2^2P)+C(h_3^2P)\\
&\leq 2C(h_2^2P)\\
&\leq \log(3/2)
\end{align}
where we used $h_3^2P\leq h_2^2P\leq1/2$. 
Therefore, if $h_2^2P\leq1/2$ we can write (by combining $\Gamma_m$ and $\Gamma_{a1}$)
\begin{align}
\max\left\{\overline{C}_\Sigma-\log\left(\frac{3}{2}\right),\frac{\overline{C}_\Sigma}{3}\right\}&\leq C_g\leq\overline{C}_\Sigma.
\end{align}

\subsection{Case $h_2^2P>1/2$}
We call the gap for this case $\Gamma_{a2}$. Notice that using $h_2^2P>1/2$ in $\underline{C}^{III}$ leads to
\begin{align}
C_g&\geq\underline{C}^{III}=2C\left(h_2^2P-\frac{1}{2}\right).
\end{align}

Now we bound $\Gamma_{a2}$ by bounding the difference between the upper bound $\overline{C}_{\Sigma g}$ and the lower bound $\underline{C}^{III}$. We obtain
\begin{align}
\Gamma_{a2}&= \overline{C}_{\Sigma g}-\underline{C}^{III}\\
&\leq C(h_2^2P+h_3^2P)+C((|h_2|+|h_3|)^2P)\nonumber\\
&\quad-2C\left(h_2^2P-1/2\right)\\
&\leq C(2h_2^2P)+C(4h_2^2P)-2C\left(h_2^2P-1\right)\\
&\leq 2C(2h_2^2P)+1/2-2C\left(h_2^2P-1\right)\\
&= \log\left(2+\frac{1}{h_2^2P}\right)+\frac{1}{2}\triangleq\overline{\Gamma}_{a2}
\end{align}
where we used $h_3^2\leq h_2^2$. Thus the gap is upper bounded by $\overline{\Gamma}_{a2}$ which approaches 3/2 as $P\to\infty$. Moreover, using $h_2^2P>1/2$ we have
\begin{align}
\Gamma_{a2}&\leq \frac{5}{2}.
\end{align}
As a result, for $h_2^2P>1/2$ we have
\begin{align}
\max\left\{\overline{C}_{\Sigma g}-\frac{5}{2},\frac{\overline{C}_\Sigma}{3}\right\}\leq C_g\leq \min\{\overline{C}_{\Sigma g},\overline{C}_\Sigma\}.
\end{align}

Thus, we have bound the gap between our sum-capacity upper and lower bounds by a constant independent of the channel coefficients. Notice that the multiplicative gap is important for the case of low power, especially when the additive gap becomes larger than the upper bound. Let us now consider the symmetric Y-channel, where $h_1=h_2=h_3=1$. In this case, given $P=P_r$, we can show that the gap between the upper and lower bounds is always less than 1 bit.

\subsection{Gap Calculation for the symmetric Y-Channel}

\begin{figure}[t]
\centering
\psfragscanon
\psfrag{x}[t]{$P_r$(dB)}
\psfrag{y}[b]{$P$(dB)}
\psfrag{z}[b]{$\Delta_g$}
\includegraphics[width=\columnwidth]{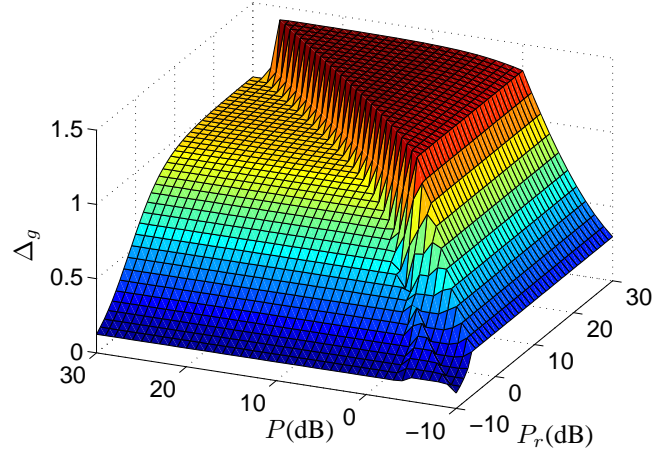}
\caption{The gap $\Delta_g$ between the upper bound and lower bound for the general symmetric Y-channel. It can be seen that the gap is always less than 1.5 bits.}
\label{Gap3Dg}
\end{figure}

In the symmetric Y-channel, $h_1=h_2=h_3=1$. In this case, we can rewrite the bounds we have in a simpler form. Starting from Corollary \ref{CSG}, we can show that the following bound holds
\begin{align}
C_g\leq\overline{C}_{cs}&\triangleq3\min\{C(P),C(P_r)\}.
\end{align}
Moreover, for the symmetric Y-channel we have the following upper bounds from Lemmas \ref{FromRelay} and \ref{GUB} respectively
\begin{align}
C_g&\leq\overline{C}_{s}\triangleq2C(2P_r)\\
C_g&\leq\overline{C}_{g}\triangleq2C(4P).
\end{align}
The following lower bounds are achievable in the symmetric Y-channel (Theorems \ref{LowerBound:DF}, \ref{LowerBound:FDF} and \ref{LowerBound:FDF_2User}) 
\begin{align}
C_g\geq\underline{C}^{i}&=\min\left\{C(3P),\frac{3}{2}C(P_r)\right\}\\
C_g\geq\underline{C}^{ii}&=2\min\left\{\left[C\left(\frac{3P}{2}-\frac{1}{2}\right)\right]^+,C(P_r)\right\}\\
C_g\geq\underline{C}^{iii}&=2\min\left\{\left[C\left(P-\frac{1}{2}\right)\right]^+,C(P_r)\right\},
\end{align}
where we used small letters in the superscript to distinguish these achievable sum rates from their counterparts in the asymmetric Y-channel. Now that we have upper and lower bounds for the sum-capacity of the symmetric Y-channel, we can upper bound the gap between them, which we denote by $\Delta_g$
\begin{align}
\label{Dg}
\Delta_g=\min\{\overline{C}_{cs},\overline{C}_s,\overline{C}_g\}-\max\{\underline{C}^i,\underline{C}^{ii},\underline{C}^{iii}\}.
\end{align}
To simplify the calculation, we assume that $P=P_r$ and bound the gap for this case. The gap for arbitrary $P$ and $P_r$ is calculated numerically and plotted in Figure \ref{Gap3Dg}.

\begin{figure}[t]
\centering
\psfragscanon
\psfrag{x}[t]{$P$(dB)}
\psfrag{y}[b]{Sum Rate}
\psfrag{UB}[bl]{{$\overline{C}$}}
\psfrag{LB}[Bl]{{$\underline{C}$}}
\psfrag{C1}{1}
\psfrag{C2}{2}
\psfrag{C3}{3}
\psfrag{C4}{4}
\includegraphics[width=\columnwidth]{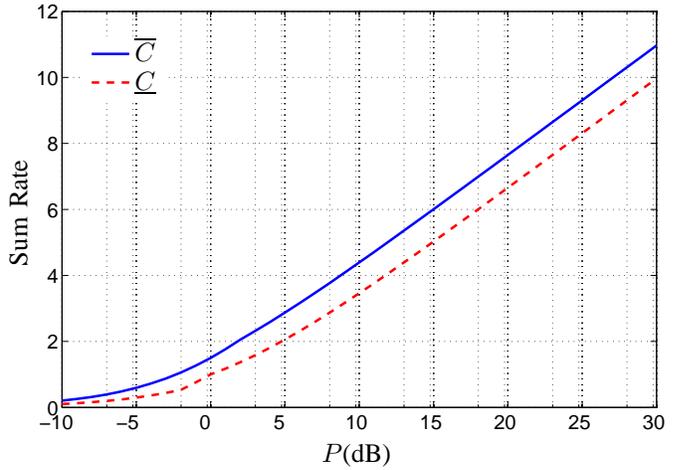}
\caption{A plot of the upper and lower bounds for the symmetric Y-channel showing the 4 cases difference cases (1-4)}
\label{Bounds_R}
\end{figure}

In the symmetric Y-channel with $P=P_r$, then we can show that $\underline{C}^{iii}\leq\underline{C}^{ii}$ and thus $\underline{C}^{iii}$ will be excluded. The upper bound $\overline{C}_g$ can also be excluded. Then, the sum-capacity is bounded as follows
\begin{align}
\underline{C}\triangleq\max\{\underline{C}^i,\underline{C}^{ii}\}\leq C_g\leq\min\{\overline{C}_{cs},\overline{C}_s\}\triangleq\overline{C}.
\end{align}
Now, bounding the difference between $\overline{C}$ and $\underline{C}$ is a simple task, and we can show that 
\begin{align}
\overline{C}-\underline{C}\leq1,
\end{align}
for any value of $P$. Figure \ref{Bounds_R} shows the upper and lower bounds for a symmetric Y-channel with $P=P_r$, where it can be seen that the gap is always less than 1 bit.

\section{Upper Bounds for the Restricted Y-channel}
\label{Section:RYC}
In this section, we impose an additional constraint on the Y-channel. That is, we consider the Y-channel with a restricted encoder (\ref{SourceEncoder_R}). Recall that the difference between the restricted Y-channel and the general one is that the transmit signals are independent in the former while they can be dependent in the later. 

The independence of the transmit signals can lead to tighter upper bound. Namely, the upper bound in Theorem \ref{SRUBG} can be tightened leading to a smaller gap to the lower bound. We start with the following lemma. 

\begin{lemma}
\label{RUB}
The achievable rates in the restricted Y-channel must satisfy
\begin{align}
R_{kj}+R_{lj}+R_{kl}&\leq C(h_k^2P+h_l^2P)
\end{align}
for all distinct $j,k,l\in\{1,2,3\}$.
\end{lemma}
\begin{proof}
We use a genie aided approach to bound the sum of three rates such as $R_{21}+R_{31}+R_{23}$ by giving $(Y_r^n,m_{32})$ and $(Y_r^n,m_{21},m_{12},m_{13})$ as additional information to receiver 1 and 3 respectively. Details are given in Appendix \ref{ProofRestricted}.
\end{proof}
Combining Lemma \ref{FromRelay} and \ref{RUB} we get for the restricted Y-channel
\begin{align}
&R_{kj}+R_{lj}+R_{kl}\nonumber\\
\label{3BoundR}
&\hspace{1cm}\leq \min\left\{C(h_j^2P_r+h_l^2P_r),C(h_k^2P+h_l^2P)\right\}
\end{align}
from which we have the following theorem.
\begin{theorem}
\label{SRUBR}
The sum-capacity of the restricted Y-channel with $P=P_r$ is upper bounded by $\overline{C}_{\Sigma r}$, i.e.
\begin{align}
\label{RSCUB}
C_r\leq\overline{C}_{\Sigma r}&=2C(h_2^2P+h_3^2P).
\end{align}                                                                                                                                                     
\end{theorem}
\begin{proof}
By evaluating (\ref{3BoundR}) for $(j,k,l)=(1,2,3)$, and for $(j,k,l)=(2,1,3)$ and  adding the two obtained bounds, we obtain the desired result.
\end{proof}

\subsection{Gap Calculation}
Keep in mind that all upper bounds for the general Y-channel continue to hold for the restricted one. This is true since $C_r\leq C_g$. However, we need not to consider $\overline{C}_{\Sigma g}$ (Theorem \ref{SRUBG}) since $\overline{C}_{\Sigma r}$ in (\ref{RSCUB}) is clearly tighter than $\overline{C}_{\Sigma g}$. Moreover, the lower bounds also hold since all achievable schemes considered above have independent transmit signals. 

While all calculated gaps hold true, the gap $\Gamma_{a2}$ can be made smaller by using $\overline{C}_{\Sigma r}$. We denote this gap for $h_2^2P>1/2$ by $\Gamma_{a2}^r$ and we bound it as follows
\begin{align}
\Gamma_{a2}^r&=\overline{C}_{\Sigma r}-\underline{C}^{III}\\
&\leq2C(2h_2^2P)-2C\left(h_2^2P-1\right)\\
&= \log\left(2+\frac{1}{h_2^2P}\right)\triangleq \overline{\Gamma}_{a2}^{r},
\end{align}
where we used $h_3^2\leq h_2^2$ (\ref{Ordering}). Notice that $\overline{\Gamma}_{2a}^{r}\to1$ as $P\to\infty$ (while $\overline{\Gamma}_{2a}\to3/2$) and using $h_2^2P>1/2$
\begin{align}
\Gamma_{2a}^{r}&\leq\overline{\Gamma}_{2a}^{r}\leq2,
\end{align}
instead of 5/2. As a result, for $h_2^2P\leq1/2$ we have
\begin{align}
\max\left\{\overline{C}_\Sigma-\log\left(\frac{3}{2}\right),\frac{\overline{C}_\Sigma}{3}\right\}&\leq C_r\leq\overline{C}_\Sigma.
\end{align}
and for $h_2^2P>1/2$ we have
\begin{align}
\max\left\{\overline{C}_{\Sigma r}-2,\frac{\overline{C}_\Sigma}{3}\right\}\leq C_r\leq \min\{\overline{C}_{\Sigma r},\overline{C}_\Sigma\}.
\end{align}

For the symmetric restricted Y-channel with $P=P_r$, the same gap of 1 bit holds as that in the asymmetric one.

\section{Summary}
\label{Summary}
We have studied the Y-channel, a system with three users and one relay where each user sends 2 messages, one to each other user via the relay. The users do not hear each other's transmission and hence the relay is essential for the communication. We studied the sum-capacity of the Y-channel by giving sum-capacity upper and lower bounds. We considered two variants: the restricted case where the transmit signal is not allowed to depend on previously received symbols, and the general case where the transmit signal is allowed to depend on previously received symbols. These bounds are derived for the Y-channel with different channel gains. The gap between the bounds is evaluated for the case of equal power at the relay and the other nodes and we have shown that this gap is less than a constant independent of the channel coefficients for both the general and the restricted setup. Hence, we characterized the sum capacity within a constant gap. For the symmetric Y-channel, where channel gains between all users and the relay are equal, we characterized the sum-capacity within one bit.

\bibliography{/home/chaaban/tex/myBib}

\begin{appendices}

\section{Proof of Corollary \ref{CSG}}
\label{CSGProof}
From the first cut-set bound (\ref{CS1}), we have
\begin{align}
R_{jk}+R_{jl}&\leq I(X_j;Y_r|X_k,X_l,X_r)\\
&=h(Y_r|X_k,X_l,X_r)-h(Z_r)\\
&\leq h(h_jX_j+Z_r)-h(Z_r)\\
\label{FCSB1}
&\leq C(h_j^2P),
\end{align}
and
\begin{align}
R_{jk}+R_{jl}&\leq I(X_r;Y_k,Y_l|X_k,X_l)\\
&=h(Y_k,Y_l|X_k,X_l)-h(Y_k,Y_l|X_k,X_l,X_r)\\
&\leq h(Y_k,Y_l)-h(Z_k,Z_l)\\
\label{FCSB2}
&\leq C(h_k^2P_r+h_l^2P_r).
\end{align}
From (\ref{FCSB1}) and (\ref{FCSB2}) we obtain (\ref{CSG1}). Using (\ref{CS2}), we have
\begin{align}
R_{jl}+R_{kl}&\leq I(X_j,X_k;Y_r|X_l,X_r)\\
&= h(Y_r|X_l,X_r)-h(Y_r|X_l,X_r,X_j,X_k)\\
&\leq h(h_jX_j+h_kX_k+Z_r)-h(Z_r)\\
&\leq C(h_j^2P+h_k^2P+2h_jh_k\rho_{jk}P)\\
\label{SCSB1}
&\leq C((|h_j|+|h_k|)^2P)
\end{align}
where $\rho_{jk}=\mathbb{E}[X_jX_k]/P\in[-1,1]$, and
\begin{align}
R_{jl}+R_{kl}&\leq I(X_r;Y_l|X_l)\\
&= h(Y_l|X_l)-h(Y_l|X_l,X_r)\\
&\leq h(Y_l)-h(Z_l)\\
\label{SCSB2}
&\leq C(h_l^2P_r).
\end{align}
From (\ref{SCSB1}) and (\ref{SCSB2}) we obtain (\ref{CSG2}).

\section{Proof of Lemma \ref{FromRelay}}
\label{ProofFromRelay}
\begin{figure}[t]
\centering
\includegraphics[width=0.8\columnwidth]{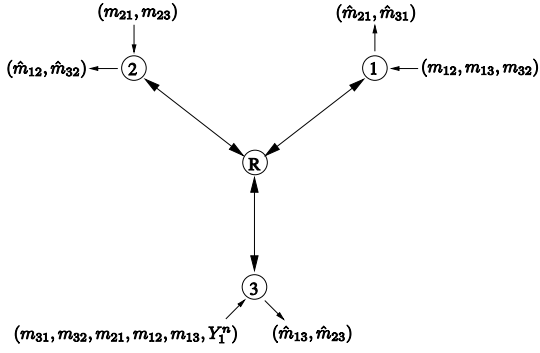}
\caption{The Y-channel with side information}
\label{GAYC_FR}
\end{figure}

Starting from Fano's inequality, we have
\begin{align}
\label{Fano1}
n(R_{21}+R_{31})&\leq I(m_{21},m_{31};Y_1^n,m_{12},m_{13})+n\epsilon_{1n}\\
\label{Fano2}
nR_{23}&\leq I(m_{23};Y_3^n,m_{31},m_{32})+n\epsilon_{2n},
\end{align}
where $\epsilon_{1n},\epsilon_{2n}\to0$ as $n\to\infty$. We give $m_{32}$ to receiver 1, and $(Y_1^n,m_{21},m_{12},m_{13})$ to receiver 3 as additional information as shown in Figure \ref{GAYC_FR} to obtain
\begin{align}
n(R_{21}+R_{31}-\epsilon_{1n})&\leq I(m_{21},m_{31};Y_1^n,m_{12},m_{13})\nonumber\\
&\leq I(m_{21},m_{31};Y_1^n,m_{12},m_{13},m_{32})\nonumber\\
\label{FP}
&= I(m_{21},m_{31};Y_1^n|m_{12},m_{13},m_{32}),
\end{align}
and
\begin{align}
n(R_{23}-\epsilon_{2n})&\leq I(m_{23};Y_3^n,m_{31},m_{32})\nonumber\\
&\leq I(m_{23};Y_3^n,m_{31},m_{32},Y_1^n,m_{21},m_{12},m_{13})\nonumber\\
&= I(m_{23};Y_1^n|m_{31},m_{32},m_{21},m_{12},m_{13})\nonumber\\
\label{SP}
&\quad+I(m_{23};Y_3^n|m_{31},m_{32},m_{21},m_{12},m_{13},Y_1^n).
\end{align}
where (\ref{FP}) and (\ref{SP}) follow by using the chain rule and from the independence of the messages. Adding (\ref{FP}) and (\ref{SP}) and using the chain rule and the non-negativity of mutual information, we get
\begin{align*}
&n(R_{21}+R_{31}+R_{23}-\epsilon_{n})\nonumber\\
&\leq I(m_{21},m_{31},m_{23};Y_1^n|m_{12},m_{13},m_{32})\nonumber\\
&\quad+I(m_{23};Y_3^n|m_{31},m_{32},m_{21},m_{12},m_{13},Y_1^n)\\
&\leq I(m_{21},m_{31},m_{23},X_r^n;Y_1^n|m_{12},m_{13},m_{32})\nonumber\\
&\quad+I(m_{23},X_r^n;Y_3^n|m_{31},m_{32},m_{21},m_{12},m_{13},Y_1^n)
\end{align*}
We continue
\begin{align*}
&n(R_{21}+R_{31}+R_{23}-\epsilon_{n})\nonumber\\
&\stackrel{(a)}{\leq} h(Y_1^n)-h(Y_1^n|X_r^n)+h(Y_3^n|Y_1^n)-h(Y_3^n|Y_1^n,X_r^n)\\
&= h(Y_1^n,Y_3^n)-h(Z_1^n,Z_3^n)\\
&\stackrel{(b)}{=} \sum_{i=1}^n\left[h(Y_{1i},Y_{3i}|Y_1^{i-1},Y_3^{i-1})-h(Z_{1i},Z_{3i})\right]\\
&\stackrel{(c)}{\leq} \sum_{i=1}^n\left[h(Y_{1i},Y_{3i})-h(Z_{1i},Z_{3i})\right]\\
&= \sum_{i=1}^n\left[h(Y_{1i},Y_{3i})\right]-n\log(2\pi e)\\
&\stackrel{(d)}{\leq} \frac{1}{2}\sum_{i=1}^n\log(1+(h_1^2+h_3^2)P_{ri})\\
&\stackrel{(e)}{\leq} \frac{n}{2}\log(1+(h_1^2+h_3^2)P_r),
\end{align*}
where $\epsilon_{n}=\epsilon_{1n}+\epsilon_{2n}\to0$ as $n\to\infty$ and
\begin{itemize}
\item[$(a)$] follows since conditioning does not increase entropy and since $Y_1^n$ and $Y_3^n$ are independent of all messages given $X_r^n$,
\item[$(b)$] follows since the noises $Z_1$ and $Z_3$ are i.i.d.
\item[$(c)$] follows since conditioning does not increase entropy,
\item[$(d)$] follows since the Gaussian distribution maximizes the differential entropy under a covariance constraint, and
\item[$(e)$] follows by using Jensen's inequality.
\end{itemize}
Thus,
\begin{align*}
R_{21}+R_{31}+R_{23}\leq C((h_1^2+h_3^2)P_r).
\end{align*}
In a similar way, we can obtain the other bounds and this completes the proof.

\section{Proof of Lemma \ref{GUB}}
\label{GeneralProof}

We start from Fano's inequality, 
\begin{align}
n(R_{21}+R_{31})&\leq I(m_{21},m_{31};Y_1^n,m_{12},m_{13})+n\epsilon_{1n}\\
nR_{23}&\leq I(m_{23};Y_3^n,m_{31},m_{32})+n\epsilon_{2n},
\end{align}
and proceed as follows
\begin{align}
n(R_{21}+R_{31}-\epsilon_{1n})&\leq I(m_{21},m_{31};Y_1^n,m_{12},m_{13})\nonumber\\
&\leq I(m_{21},m_{31};Y_1^n,m_{12},m_{13},Y_r^n,m_{32})\nonumber\\
&= I(m_{21},m_{31};m_{12},m_{13},m_{32})\nonumber\\
&\quad+I(m_{21},m_{31};Y_r^n|m_{12},m_{13},m_{32})\nonumber\\
&\quad+I(m_{21},m_{31};Y_1^n|m_{12},m_{13},m_{32},Y_r^n)\nonumber\\
\label{CFP}
&= I(m_{21},m_{31};Y_r^n|m_{12},m_{13},m_{32}),
\end{align}
where (\ref{CFP}) follows since the messages $m_{ij}$ are all independent, and from the Markov chain $(m_{21},m_{31})\to Y_r^n\to Y_1^n$.
\begin{align}
n(R_{23}-\epsilon_{2n})&\leq I(m_{23};Y_3^n,m_{31},m_{32})\nonumber\\
&\leq I(m_{23};Y_3^n,m_{31},m_{32},Y_r^n,m_{21},m_{12},m_{13})\nonumber\\
&= I(m_{23};m_{31},m_{32},m_{21},m_{12},m_{13})\nonumber\\
&\quad +I(m_{23};Y_r^n|m_{31},m_{32},m_{21},m_{12},m_{13})\nonumber\\
&\quad +I(m_{23};Y_3^n|m_{31},m_{32},m_{21},m_{12},m_{13},Y_r^n)\nonumber\\
\label{CSP}
&= I(m_{23};Y_r^n|m_{31},m_{32},m_{21},m_{12},m_{13}),
\end{align}
where (\ref{CSP}) follows since the messages $m_{ij}$ are all independent, and from the Markov chain $m_{23}\to Y_r^n\to Y_3^n$. Adding these inequalities, we obtain
\begin{align}
&\hspace{-1cm}n(R_{21}+R_{31}+R_{23}-\epsilon_{n})\nonumber\\
&\quad\leq I(m_{21},m_{31};Y_r^n|m_{12},m_{13},m_{32})\nonumber\\
&\quad\quad+I(m_{23};Y_r^n|m_{31},m_{32},m_{21},m_{12},m_{13})\nonumber\\
\label{Multi_Letter_UB}
&\quad=I(m_{21},m_{31},m_{23};Y_r^n|m_{12},m_{13},m_{32}),
\end{align}
where $\epsilon_{n}=\epsilon_{1n}+\epsilon_{2n}$. In what follows, we will use the following notation 
\begin{align*}
\vec{Z}^n&\triangleq(Z_1^n,Z_2^n,Z_3^n),\\
\vec{Y}^n&\triangleq(Y_1^n,Y_2^n,Y_3^n).
\end{align*}
We proceed as follows
\begin{align*}
&n(R_{21}+R_{31}+R_{23}-\epsilon_{n})\\
&\leq I(m_{21},m_{31},m_{23};Y_r^n|m_{12},m_{13},m_{32})\\
&\leq I(m_{21},m_{31},m_{23};Y_r^n,\vec{Z}^n|m_{12},m_{13},m_{32})\\
&\stackrel{(a)}{=} I(m_{21},m_{31},m_{23};Y_r^n|m_{12},m_{13},m_{32},\vec{Z}^n)
\end{align*}
where $(a)$ follows since the messages and $\underline{Z}^n$ are independent. Then
\begin{align*}
&n(R_{21}+R_{31}+R_{23}-\epsilon_{n})\\
&\leq \sum_{i=1}^n I(m_{21},m_{31},m_{23};Y_{ri}|m_{12},m_{13},m_{32},\vec{Z}^n,Y_{r}^{i-1})\\
&\stackrel{(b)}{=} \sum_{i=1}^n I(m_{21},m_{31},m_{23};Y_{ri}|m_{12},m_{13},m_{32},\vec{Z}^n,Y_{r}^{i-1},X_r^i)\\
&\stackrel{(c)}{=} \sum_{i=1}^n h(Y_{ri}|m_{12},m_{13},m_{32},\vec{Z}^n,Y_{r}^{i-1},X_r^i,\vec{Y}^{i},X_{1i})\\
&\quad-\sum_{i=1}^n h(Y_{ri}|\vec{m},\vec{Z}^n,Y_{r}^{i-1},X_r^i,\vec{Y}^{i},X_{1i},X_{2i},X_{3i})\\
&\stackrel{(d)}{\leq} \sum_{i=1}^n \left[h(Y_{ri}|X_{1i})-h(Y_{ri}|X_{1i},X_{2i},X_{3i})\right]\\
&\leq \sum_{i=1}^n \left[h(h_2X_{2i}+h_3X_{3i}+Z_{ri})-h(Z_{ri})\right],
\end{align*}
where
\begin{itemize}
\item[$(b)$] follows since $X_r^i=f_r(Y_r^{i-1})$ (\ref{RelayEncoder}),
\item[$(c)$] follows since $Y_j^i=h_jX_r^i+Z_j^i$ with $j\in\{1,2,3\}$ (\ref{ReceivedSignal}) and since in the general Y-channel (\ref{SourceEncoder}) 
\begin{align}
X_{1i}&=f_1(m_{12},m_{13},Y_1^{i-1}),\\
X_{2i}&=f_2(m_{21},m_{23},Y_2^{i-1}),\\
X_{3i}&=f_3(m_{31},m_{32},Y_3^{i-1}), \text{ and}
\end{align}
\item[$(d)$] follows since conditioning does not increase entropy, and since the channel is memoryless.
\end{itemize}
This upper bound is maximized by Gaussian $X_{2i}$ and $X_{3i}$ since the circularly symmetric Gaussian distribution maximizes the differential entropy under a covariance constraint. Since in the general Y-channel, the transmit symbols are allowed to depend on past received symbols, the transmit symbols at different users can be correlated. Let $(X_{2i},X_{3i})$ be a Gaussian vector with zero mean and covariance matrix 
\begin{equation}
\Sigma(X_{2i},X_{3i})=\left(\begin{array}{cc}
P_{2i} &\rho_{23}\sqrt{P_{2i}P_{3i}}\\
\rho_{23}\sqrt{P_{2i}P_{3i}} & P_{3i}
\end{array}\right), 
\end{equation}
with $\rho_{23}\in[-1,1]$. Then, $\mathbb{E}[(h_2X_{2i}+h_3X_{3i})^2]=h_2^2P_{2i}+h_3^2P_{3i}+2h_2h_3\rho_{23}\sqrt{P_{2i}P_{3i}}$. Therefore
\begin{align*}
&n(R_{21}+R_{31}+R_{23}-\epsilon_{n})\\
&\leq \sum_{i=1}^n h(h_2X_{2i}+h_3X_{3i}+Z_{ri})-h(Z_{ri})\\
&\leq \sum_{i=1}^n \frac{1}{2}\log\left(1+h_2^2P_{2i}+h_3^2P_{3i}+2h_2h_3\rho_{23}\sqrt{P_{2i}P_{3i}}\right)\\
&\stackrel{(e)}{\leq} \sum_{i=1}^n \frac{1}{2}\log\left(1+\left(\sqrt{h_2^2P_{2i}}+\sqrt{h_3^2P_{3i}}\right)^2\right)\\
&\stackrel{(f)}{\leq} \frac{n}{2} \log\left(1+(|h_2|+|h_3|)^2P\right),
\end{align*}
where $(e)$ follows by using $h_2h_3\rho_{23}\leq|h_2||h_3|$ since $\rho_{23}$ with 1, and $(f)$ follows by using Jensen's inequality on a function that can be proved to be concave\footnote{Since the function $f(x)=\log(1+x)$ is concave and non-decreasing, $f((\sqrt{x}+\sqrt{y})^2)$ is concave if the function $g(x)=(\sqrt{x}+\sqrt{y})^2$ is concave as well. Thus it is sufficient to show that $(\sqrt{x}+\sqrt{y})^2$ is concave which can be shown to be true by checking its Hessian for example.}. Letting $n\to\infty$, we obtain
\begin{align}
\label{BG1}
R_{21}+R_{31}+R_{23}\leq C((|h_2|+|h_3|)^2P).
\end{align}
The other bounds can be obtained in a similar way, and this ends the proof.

\section{Solution of the linear program in (\ref{OptProb})}
\label{LPMax}
Let us use the following notation $A=C(h_1^2P_r)$, $B=C(h_2^2P_r)$, $C=C(h_1^2P_r)$,
\begin{align}
\label{x}
x&=R_{21}+R_{23}\\
y&=R_{31}+R_{32}\\
\label{z}
z&=R_{12}+R_{13}. 
\end{align}
Notice from (\ref{Ordering}) that $A\geq B\geq C$. We then solve the following linear program
\begin{align}
\label{LP2}
\text{maximize}\quad &x+y+z\\
\text{subject to}\quad & x,y,z\geq0\nonumber\\
&x+y\leq A\nonumber\\
&y+z\leq B\nonumber\\
&z+x\leq C.\nonumber
\end{align}

\begin{figure}
\centering
\subfigure[Feasible region of (\ref{LP2}) when $A<B+C$.]{
\psfragscanon
\psfrag{x}[t]{$x$}
\psfrag{y}[b]{$y$}
\psfrag{z}[l]{$z$}
\psfrag{N}[l]{$N$}
\includegraphics[width=0.4\columnwidth]{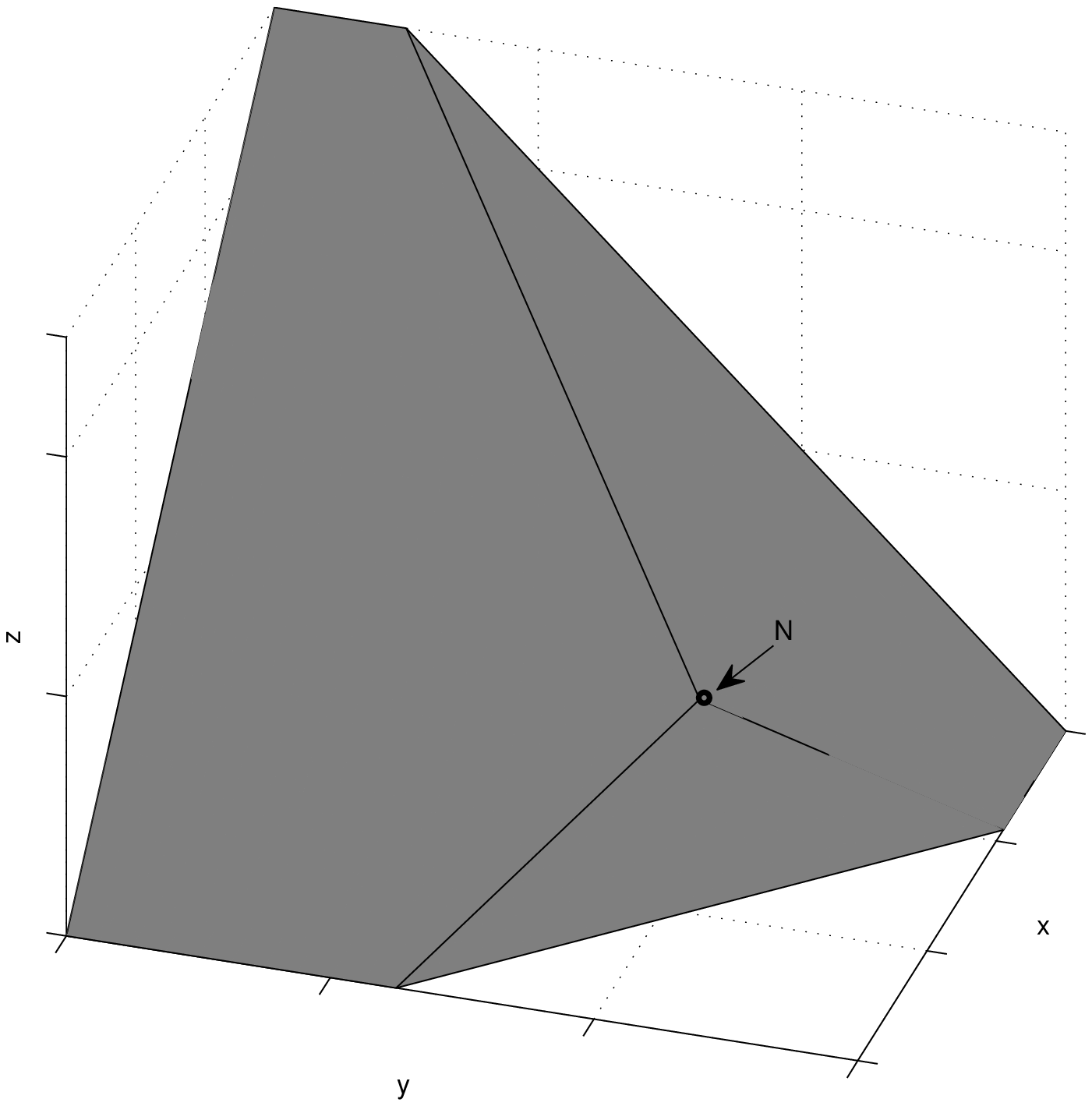}
\label{LPFig1}
}
\hspace{0.5cm}
\subfigure[Feasible region of (\ref{LP2}) when $A\geq B+C$.]{
\psfragscanon
\psfrag{x}[t]{$x$}
\psfrag{y}[b]{$y$}
\psfrag{z}[l]{$z$}
\psfrag{M}[b]{$M$}
\includegraphics[width=0.4\columnwidth]{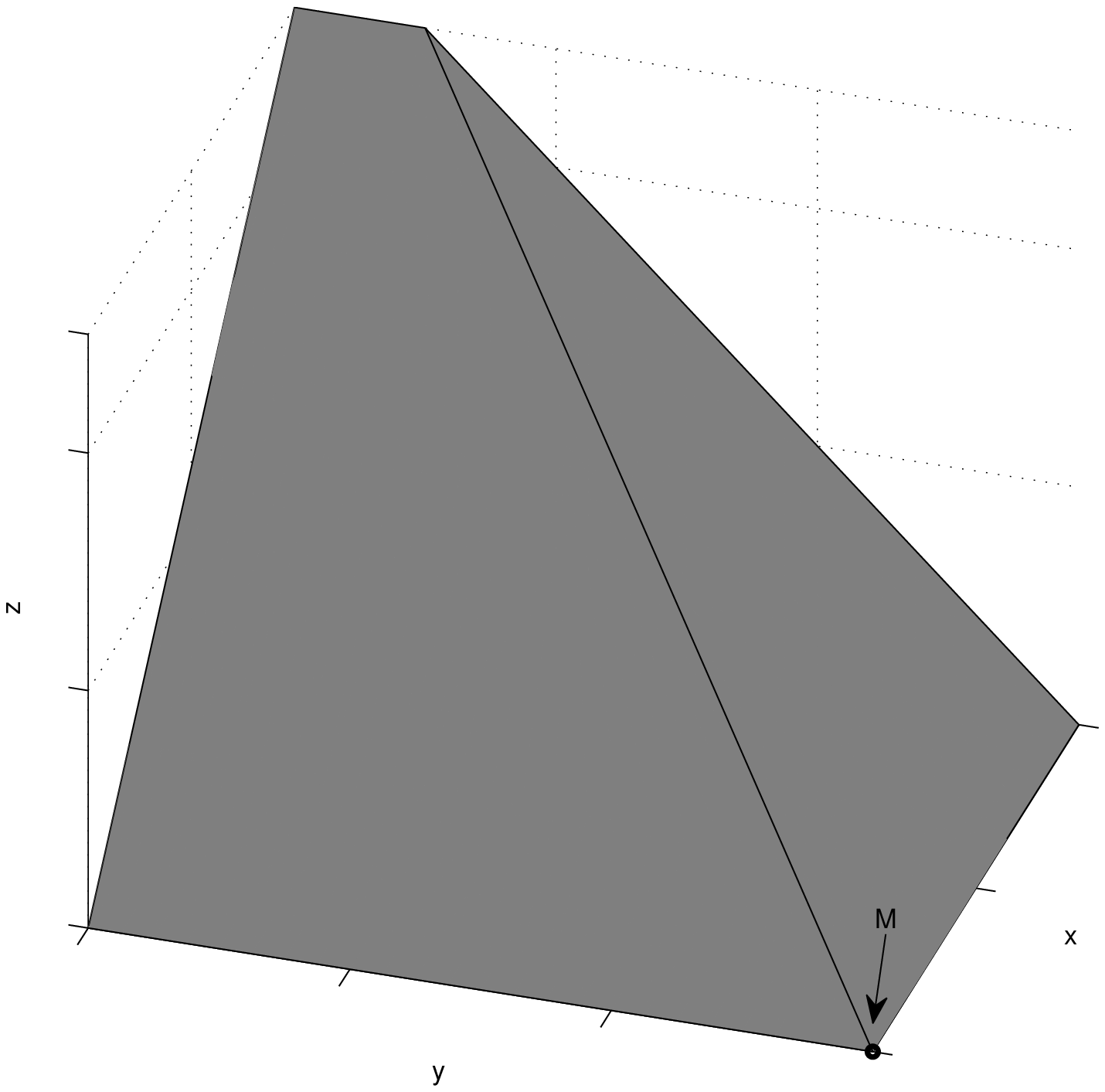}
\label{LPFig2}
}
\caption{Sets of feasible points of problem \eqref{LP2}.}
\end{figure}

The conditions $x,y,z\geq0$ are less stringent than $R_{jk}\geq0,\forall j,k\in\{1,2,3\},\ j\neq k$, hence the solution of (\ref{LP2}) is not smaller than that of (\ref{OptProb}). Moreover, for every feasible point $(x,y,z)$ in (\ref{LP2}), there exist $R_{jk}\geq0$ satisfying (\ref{x})-(\ref{z}). Therefore, the solution of (\ref{LP2}) is equal to the solution of (\ref{OptProb}), thus solving this linear program leads to the solution of the original problem in (\ref{OptProb}). The feasible set in (\ref{LP2}) forms a polyhedron that can have two different forms: 
\begin{itemize}
\item{(a)} if $A<B+C$ then the feasible set is the polyhedron in Figure \ref{LPFig1}, 
\item{(b)} if $A\geq B+C$, then the feasible set is the polyhedron in Figure \ref{LPFig2}.
\end{itemize}

Using the simplex method, the point that maximizes $x+y+z$ is the corner point $$N=\frac{1}{2}(A-B+C,A+B-C,-A+B+C),$$ in case (a), and is the corner point $M=(C,B,0)$ in case (b).

Therefore, the solution of (\ref{LP2}) is
\begin{align}
\left\{
\begin{array}{lr}
\frac{1}{2}(A+B+C)& \text{if } A< B+C\\
B+C& \text{otherwise}
\end{array}\right.
\end{align}
which can also be written as
\begin{align}
\min\left\{\frac{1}{2}(A+B+C),B+C\right\}.
\end{align}

\section{Proof of Lemma \ref{RUB}}
\label{ProofRestricted}
We start from inequality (\ref{Multi_Letter_UB}) which also holds for the restricted Y-channel. Now we can write
\begin{align*}
&n(R_{21}+R_{31}+R_{23}-\epsilon_{n})\\
&\leq I(m_{21},m_{31},m_{23};Y_r^n|m_{12},m_{13},m_{32})\\
&\stackrel{(a)}{=} h(Y_r^n|m_{12},m_{13},m_{32},X_1^n)-h(Y_r^n|\vec{m},X_1^n,X_2^n,X_3^n)\\
&\stackrel{(b)}{\leq} h(h_2X_2^n+h_3X_3^n+Z_r^n)-h(Z_r^n)\\
&\stackrel{(c)}{\leq} \sum_{i=1}^n h(h_2X_{2i}+h_3X_{3i}+Z_{ri})-\frac{n}{2}\log(2\pi e)\\
&\stackrel{(d)}{\leq} \sum_{i=1}^n \frac{1}{2}\log(1+h_2^2P_{2i}+h_3^2P_{3i})\\
&\stackrel{(e)}{\leq} \frac{n}{2}\log(1+h_2^2P+h_3^2P),
\end{align*}
where
\begin{itemize}
\item[$(a)$] follows since the Y-channel is restricted, i.e. $X^n_j=f_1(m_{jk},m_{jl})$, $\{j,k,l\}=\{1,2,3\}$ (\ref{SourceEncoder}), and by denoting $(m_{12},m_{13},m_{32},m_{21},m_{31},m_{23})$ by $\vec{m}$,
\item[$(b)$] follows since conditioning does not increase entropy, and since $Z_r^n$ is independent of the messages and the transmit signals,
\item[$(c)$] follows by using the chain rule and the fact that conditioning does not increase entropy,
\item[$(d)$] follows since the Gaussian distribution maximizes the differential entropy under a covariance constraint, and since the channel is restricted, thus the signals $X_{2i}$ and $X_{3i}$ are not correlated, and
\item[$(e)$] follows by using Jensen's inequality on a function that can be proved to be concave.
\end{itemize}
Letting $n\to\infty$ we obtain
\begin{align}
\label{B1}
R_{21}+R_{31}+R_{23}\leq C((h_2^2+h_3^2)P).
\end{align}
Similarly we can obtain the other bounds and this completes the proof.

\end{appendices}

\end{document}